\documentstyle[prd,aps,epsfig]{revtex}

\begin{document}

\draft

\title{Counting defects in an instantaneous quench}

\author{
D. Ibaceta\thanks{Electronic address: {\tt ivan@df.uba.ar}} 
and 
E. Calzetta\thanks{Electronic address: {\tt calzetta@df.uba.ar}}}
\address{Department of Physics and IAFE, University of Buenos Aires, Argentina}
\date{\today}
\maketitle

\begin{abstract}
We consider the formation of defects in a non equilibrium second order phase 
transition induced by an instantaneous quench to zero temperature in a type 
II superconductor.
We perform a full non-linear simulation where we follow the evolution in time 
of the local order parameter field.
We determine how far into the phase transition theoretical estimates of the 
defect density based on the gaussian approximation yield a reliable prediction 
for the actual density.
We also characterize quantitatively some aspects of the out of equilibrium 
phase transition.
\end{abstract}


\section{Introduction}

The objective of this paper is to study the formation of defects in a non
equilibrium second order phase transition by means of a numerical solution
of the full dynamical equations, and to compare the results with theoretical
predictions to be found in the literature 
\cite{Halperin,Liu-Mazenko,Karra-Rivers}.

Topological defects are a common occurrence in symmetry broken field
theories \cite{Nash-Sen}, with far reaching consequences in condensed matter
physics \cite{cmp}, particle physics \cite{Skyrme,Huang} and cosmology \cite
{Vilenkin}. The equilibrium structure of defects is deeply rooted in the
topological aspects of the theory and it is well understood. The dynamical
formation of defects in the process of a non equilibrium phase transition,
on the other hand, only recently has been the subject of a systematic
analysis.

The issue at stake is how many defects are to be formed, as a function of
both the dynamics of the system and the macroscopic parameters
characterizing the transition, such as cooling rates. A simple, order of
magnitude estimate is based on the observation that once the system is cold
enough, defects will be unable to either form or disappear through thermal
activation, so they will be essentially frozen into existence. This happens
at the so-called Ginzburg temperature, and leads to the prediction that the
typical distance between defects is of the order of magnitude of the
correlation length at this temperature \cite{Kibble}.

This picture of defect dynamics has been criticized by Zurek as downplaying
the non equilibrium features of the process \cite{Zurek,Yates-Zurek}. According to this
author, the freezing of defects occurs at a much higher temperature, when
the relaxation time of the system becomes large and the system effectively
decouples from its environment. Zurek's arguments usually lead to higher
defect densities than previously though, a prediction confirmed by some
experiments \cite{liqcrys,experiments} (see however \cite{He4}). They are supported also by numerical
simulations in one and two dimensional Ginzburg-Landau systems with broken
global symmetries \cite{numsim}.

Complementary to the search for a qualitative understanding of defect
formation, some authors have attempted to derive the density of defects from
a first principles description of the dynamics. As a matter of fact, there
is a rigorous result linking the density of defects to the equal time
correlation function for the order parameter, valid whenever the field
probability distribution is Gaussian \cite{Halperin,Liu-Mazenko}. Under the
Gaussian approximation, then, finding the density of defects is reduced to
solving the dynamics for the correlation function \cite{Karra-Rivers}. This
approach, which has been recently used to show the validity of Zurek's
estimates in a quantum field theoretic model \cite{Greg}, is usually though
to be correct early in the development of the phase transition 
\cite{Karra-Rivers 2}.

In order to evaluate how reliable this kind of argument really is,
therefore, we must know how far in the phase transition the Gaussian
approximation may be trusted. Since the development of the phase transition
is an essentially nonlinear process, we must expect that non Gaussian
correlations will be created by the dynamics itself, even if suppressed in
the initial conditions. There will be a competition between the
characteristic growth time due to the spinodal instability 
\cite{spinodal,Boyanovsky}, and a variety of dynamical times describing the
building of correlations through non linear interaction of fluctuations;
however, the usual Gaussian models do not give us a clue about what the
latter might be.

The objective of this paper is to give a tentative answer to the question of
the reliability of defect density estimates based on the Gaussian
approximation, by presenting a fully nonlinear simulation of a phase
transition where we have measured the Gaussianity of the order parameter,
the Gaussian prediction for the density of defects, and the actual density,
as functions of time. Of course, as long as the field is actually Gaussian,
the rigorous analytic prediction for the defect density is validated. The
nontrivial issue is how long the field probability density remains Gaussian,
and whether the Gaussian prediction continues to hold beyond that point.
Concretely, we find that even when gaussianity of the ensemble ceases to
hold, the prediction remains valid, to break down about the time defects are
definitely formed.

Our simulations follow the unfolding of an instantaneous quench to zero 
temperature in a two dimensional type II superconductor, described by the 
time-dependent Landau-Ginzburg equations derived by Gor'kov and Eliashberg
\cite{GyE} (see also \cite{Tinkham,gorkov,Hu-Thompson,Kato}). This model is a 
scalar field theory interacting with a $U(1)$ gauge field 
(see \cite{Yates-Zurek}) but with the presence of a normal current, besides 
the supercurrent, and a first order (rather than second one) equation for the 
vector potential ${\bf A}$ (see below).

In order to perform the simulations, we have discretized the model and
placed it on a square lattice, with periodic boundary conditions. Our
discrete model still has gauge symmetry, which is preserved by the
evolution. We have further considered a variety of lattice sizes and initial
conditions, thus making sure that our results truly reflect the physics of
the system. We have identified defects by measuring the winding number of
the field around each lattice plaquette, thus avoiding the uncertainties
related to identifying defects from zeroes in the order parameter 
\cite{Habib}. 

We find that the evolution of a typical quench goes through three well
defined regimes. The early development is dominated by the exponential
growth of the order parameter; the different modes of the field evolve
independently, and the field remains Gaussian. In this regime, the Halperin
- Mazenko - Liu (HML) prediction for the defect density is very accurate.
This regime ends when the order parameter reaches about a tenth of its
equilibrium value.

The second regime is a transitional epoch dominated by the actual formation
of the defects. 
During this transitional epoch, the field departs from gaussianity in a 
significative way, but the HML prediction is still a good approximation 
to the actual density.
Finally, in the late time regime both the Gaussian approximation for the order
parameter and the HML prediction are unreliable.

We may therefore conclude that, as argued by Karra and Rivers 
\cite{Karra-Rivers}, the HML prediction holds early in the development of the
phase transition, being a very accurate estimate of the actual density until
the time the defects may be considered as definitely formed, which is also
the time when the order parameter reaches about a half of its equilibrium
value. It is therefore a suitable means to estimate the initial conditions
for the subsequent evolution of the defect network, as determined by defect
- defect interactions and changes in the environment, such as the expansion
of the Universe in cosmological applications \cite{Vilenkin}.
Our results validate previous analysis of non equilibrium defect formation,
such as \cite{Halperin,Liu-Mazenko,Karra-Rivers,Greg}.

The paper is organized as follows. The first part introduces the theoretical
prediction of the defect density for a gaussian quench. The second part
shows the time-dependent Ginzburg-Landau model, and the numerical details
related to its implementation. The third part describes the resulting quenches 
and the conclusions, and the paper ends with some final remarks.

\section{Theoretical prediction}

At a qualitative level, the formation of topological defects is well
understood through the symmetry breaking mechanism. For a complex scalar
field, the true ground-state manifold of the field is $S1$ and the phase of
the field at different points can be different. If eventually the winding
number of the field along a closed loop is not zero, then topological
defects are trapped inside. Because of topological considerations, isolated
defects at low temperature are stable, although defects may interact with
each other and annihilate, migrate to the boundaries of the system, or, in
cosmology, decay through gravitational radiation.

We are searching for the monopoles of one complex order parameter field 
$\Phi $ in two dimensions. This will be identified with the zeroes of the
field with non-trivial winding number. If they are located at 
${\bf x}_1,{\bf x}_2,{\bf x}_3,...$, 
we obtain for the total and topological densities 
\begin{eqnarray}
\overline{\rho }({\bf x}) &=&\sum_i\delta ({\bf x}-{\bf x}_i) \\
\rho ({\bf x}) &=&\sum_in_i\delta ({\bf x}-{\bf x}_i) \nonumber
\end{eqnarray}

Where $n_i$ is the winding number of each defect, i.e. its topological
charge.

The total density of defects is obtained through the relation between zeros
of the field and the field itself, i.e. the Jacobian \cite{Karra-Rivers}, 
\begin{equation}
\overline{n}(t)=\left\langle \overline{\rho }({\bf x})\right\rangle =\int 
{\cal D}\Phi p_t\left[ \Phi \right] \delta ^2\left[ \Phi \right] \left|
\epsilon _{jk}\partial _j\Phi _1({\bf x})\partial _k\Phi _2({\bf x})\right|
\label{2} 
\end{equation}
with $\epsilon _{12}=-\epsilon _{21}=1$ (otherwise zero) and 
$\Phi =(\Phi _1+i\Phi _2)/\sqrt{2}$, $p_t$ being the probability
density of the different field configurations.

For the Gaussian model we have $\left\langle \Phi _a({\bf x})\right\rangle
=0=\left\langle \Phi _a({\bf x})\partial _j\Phi _b({\bf x})\right\rangle $.
Assuming also that the equal-time Wightman function $\left\langle \Phi _a(%
{\bf x})\Phi _b({\bf y})\right\rangle =W_{ab}(\left| {\bf x-y}\right|
;t)=\delta _{ab}W$ $(\left| {\bf x-y}\right| ;t)$ is the only non-vanishing
correlation function and it is diagonal, then \cite{Halperin,Liu-Mazenko}: 
\begin{equation}
\overline{n}(t)=\frac 1{2\pi }(-f^{\prime \prime }(0;t))  \label{KR 0}
\end{equation}

where

\begin{equation}
f(r;t)=\frac{W(r;t)}{W(0;t)} 
\end{equation}

and derivatives are taken with respect to $r$.

To compute the r.h.s. of (\ref{KR 0}) we consider the Fourier transform of
the field, $\widetilde{\Phi }({\bf k})=\int d{\bf x\exp (}i{\bf k.x)}\Phi 
{\bf (x)}$ in order to obtain 
\begin{equation}
f^{\prime \prime }(0;t)=-\frac{\int d{\bf k}k^2\left| \widetilde{\Phi }({\bf %
k})\right| ^2}{\int d{\bf k}\left| \widetilde{\Phi }({\bf k})\right| ^2}
\label{integral kcu}
\end{equation}

Our goal is to test the range of validity of the relation (\ref{KR 0}) by
measuring both sides of this equation independently.

\section{The model}

\subsection{Theory}

The time-dependent Ginzburg-Landau equation describes the time-space
dependence of the order parameter of a superconductor \cite{GyE}. 
Normalized in the form adopted by Hu and Thompson, it reads 
\cite{Tinkham,gorkov,Hu-Thompson} 
\begin{equation}
\frac 1D\left[ \frac \partial {\partial t}+i\frac{2e}\hbar \psi \right]
\Delta +\xi (T)^{-2}\left( \left| \Delta \right| ^2-1\right) \Delta +\left[ 
\frac{{\bf \nabla }}i-\frac{2e}{\hbar c}{\bf A}\right] ^2\Delta -f({\bf r}%
,t)=0  \label{TDGL-continua}
\end{equation}

where

\begin{eqnarray}
{\bf j} &=&\sigma \left[ -{\bf \nabla }\psi -\frac 1c\frac{\partial {\bf A}}{%
\partial t}\right] +%
\mathop{\rm Re}
\left[ \Delta ^{*}\left[ \frac{{\bf \nabla }}i-\frac{2e}{\hbar c}{\bf A}%
\right] \Delta \right] \frac{\hbar c^2}{8\pi e\lambda (T)^2} \\
\rho &=&\frac{\psi -\varphi }{4\pi \lambda _{TF}^2}
\end{eqnarray}

With the Maxwell equations coupling the electromagnetic potentials to charge
and current densities, they provide the full set of evolution equations.
Here, $D$ is the normal-state diffusion constant, $\sigma $ is the
normal-state conductivity given by 
$\sigma =\frac{c^2\xi ^2}{48\pi \lambda ^2}\frac 1D$, $\rho $ and $j$ are 
the charge and current densities. $f$ is a
finite temperature random driving force; since we shall consider a quench to
zero temperature, it will eventually be set to zero (see below).
Furthermore, the order parameter is divided by its equilibrium value 
$\Delta_\infty =\pi k\sqrt{2\left( T_c^2-T^2\right) }$, 
where $T_c$ is the critical
temperature. Note that this is a temperature-dependent parametrization.

${\bf A}$ and $\varphi $ are the vector and scalar potentials respectively,
and $\psi $ is the electrochemical potential divided by the electronic
charge. Assuming $\rho =0$, that is, absence of net charge at the grid scale, 
results in $\psi =\varphi $ (see ref. \cite{GyE,Tinkham}).

This set of equations is invariant under the gauge transformation 
\begin{eqnarray}
{\bf A} &\rightarrow &{\bf A}-{\bf \nabla }\chi  \label{GAUGE-continua} \\
\varphi &\rightarrow &\varphi +\frac 1c\frac{\partial \chi }{\partial t} 
\nonumber \\
\Delta &\rightarrow &\Delta \exp \left[ -i\frac{2e}{\hbar c}\chi \right] 
\nonumber
\end{eqnarray}

From the microscopic theory we have the relationships 
\begin{equation}
\frac{4\pi \lambda (T)^2\sigma }{c^2}=\frac{\xi (T)^2}{12D}=\frac{\pi \hbar 
}{96k_BT_c}\left( 1-\frac T{T_c}\right) ^{-1}\equiv \frac{t_{GL0}}{12}\left(
1-\frac T{T_c}\right) ^{-1} 
\end{equation}
where $\xi (T)=\xi (0)\left[ 1-\frac T{T_c}\right] ^{-1/2}$ and $\lambda
(T)=\lambda (0)\left[ 1-\frac T{T_c}\right] ^{-1/2}$ are the temperature
dependent correlation (coherence) length and magnetic penetration depth
respectively, and $t_{GL0}=\frac{\pi \hbar }{8k_BT_c}$ is the characteristic
relaxation time of the uniform mode at zero temperature.

We can write our model in terms of dimensionless variables as follows 
\cite{Kato} 
\begin{eqnarray}
t &\rightarrow &t\text{ }t_0\text{ with }t_0=\frac{\pi \hbar }{96k_BT_c}=%
\frac{t_{GL0}}{12}  \label{SCALING} \\
r &\rightarrow &r\text{ }\xi \left( 0\right)  \nonumber \\
A &\rightarrow &A\text{ }\frac{\Phi _0}{2\pi \xi \left( 0\right) }\text{
with }\Phi _0=\frac{hc}{2e}  \nonumber \\
\varphi &\rightarrow &\varphi \text{ }\frac{\Phi _0}{2\pi ct_0}  \nonumber \\
j &\rightarrow &j\text{ }\frac{c\Phi _0}{8\pi ^2\xi \left( 0\right) } 
\nonumber \\
f &\rightarrow &f\text{ }\xi \left( 0\right) ^2  \nonumber \\
T &\rightarrow &T\text{ }T_c  \nonumber
\end{eqnarray}
to obtain \cite{GyE,Tinkham,gorkov,Hu-Thompson}
\begin{eqnarray}
\frac \partial {\partial t}\Delta +i\varphi \triangle &=&-\frac 1{12}\left[
\left( i{\bf \nabla }+{\bf A}\right) ^2\triangle +\left( 1-T\right) \left(
\left| \Delta \right| ^2-1\right) \Delta -f\right]  \label{TDGL-adim} \\
\frac \partial {\partial t}{\bf A+\nabla }\varphi &=&\left( 1-T\right) 
\mathop{\rm Re}
\left[ \Delta ^{*}\left( -i{\bf \nabla }-{\bf A}\right) \Delta \right]
-\kappa ^2{\bf \nabla }\times ({\bf \nabla }\times {\bf A})  \nonumber
\end{eqnarray}
where $\kappa =\lambda (T)/\xi (T)$ is the temperature independent
Ginzburg-Landau parameter which characterizes the superconductor. For a type
II superconductor we have $\kappa >1/\sqrt{2}$.
We have chosen $\kappa =\sqrt{2}$.

The gauge freedom allows us to choose $\varphi \equiv 0$.

Since we are interested in instantaneous quenches towards zero temperature,
we set $T=0$, and $f=0$ \onlinecite{notaBray}. 
Furthermore we must prepare the system in some thermal equilibrium
configuration. This means to generate a set of initial conditions
corresponding to a thermal distribution of modes \cite{Tinkham} : 
\begin{eqnarray}
\left\langle \widetilde{\Delta }\left( {\bf k}\right) \right\rangle &=&0
\label{tdm} \\
\left\langle \widetilde{\Delta }\left( {\bf k}\right) \widetilde{\Delta }%
\left( {\bf 0}\right) \right\rangle &=&\frac 1V\frac{2m^{*}}{\hbar ^2}\frac{%
k_BT}{k^2+1/\xi (T)^2}  \nonumber
\end{eqnarray}
with a cutoff when $k\approx \xi _0^{-1}$ beyond which GL theory is not valid.
Here $m^{*}=2m_e$, is the mass of the coupled electrons. In terms of
dimensionless variables, 
\begin{equation}
\left\langle \left| \widetilde{\Delta }\left( {\bf k}\right) \right|
^2\right\rangle =\frac \mu V\frac T{k^2+\xi ^{-2}}\text{ with }\mu =\frac{%
2k_BT_cm^{*}\xi \left( 0\right) }{\hbar ^2}  \label{Amplitud termica}
\end{equation}

The factor $\mu $ is clearly substance dependent and will be chosen later on.

\subsection{Implementation}

The discrete version of the gauge transformation is obtained through

\begin{eqnarray}
A_\mu ^j &\rightarrow &A_\mu ^j-\frac{{\bf (}\chi ^{j+\mu }-\chi ^j)}{a_\mu }
\label{GAUGE-discreta} \\
\Delta ^j &\rightarrow &\Delta ^j\exp \left[ -i\chi ^j\right]  \nonumber
\end{eqnarray}
where $\mu $ stands for a direction and $j$ for a site in the lattice. In
order to obtain a discrete version of (\ref{TDGL-adim}), invariant under (%
\ref{GAUGE-discreta}) we employ the usual link variables technique from 
Lattice QCD \cite{Huang} as in \cite{Kato}:

\begin{equation}
U_\mu ^{r_1r_2}=\exp \left[ -i\int\limits_{r_1}^{r_2}A_\mu d\mu \right]
\rightarrow \text{discretized }U_\mu ^{j,\text{ }j+\mu }=\exp \left[ -iA_\mu
^ja_\mu \right] 
\end{equation}

The differential operators become 
\begin{eqnarray}
\left[ \frac 1i\frac \partial {\partial x_\mu }-A_\mu \right] \Delta 
&\rightarrow &-i\frac{U_\mu ^{j+\mu ,\text{ }j}\Delta ^{j+\mu }-\Delta ^j}{%
a_\mu } \\
\left[ \frac 1i\frac \partial {\partial x_\mu }-A_\mu \right] ^2\Delta 
&\rightarrow &\frac{U_\mu ^{j+\mu ,\text{ }j}\Delta ^{j+\mu }-2\Delta
^j+U_\mu ^{j-\mu ,\text{ }j}\Delta ^{j-\mu }}{a_\mu ^2}
\end{eqnarray}
and the finite difference equations to solve are ( $\Delta ^j=\rho
^je^{i\theta ^j}$ ): 
\begin{eqnarray}
\mathop{\rm Re}
\stackrel{.}{\Delta }^j &=&\frac 1{12}\left[ \left( \rho ^{j+x}\cos
(-A_x^ja_x+\theta ^{j+x})-2%
\mathop{\rm Re}
\Delta ^j+\rho ^{j-x}\cos (A_x^{j-x}a_x+\theta ^{j-x})\right) \frac 1{a_x^2}%
+...\right.   \label{discreta orden} \\
&&\left. -(1-T)(\rho ^{j2}-1)%
\mathop{\rm Re}
\Delta ^j\right]   \nonumber
\end{eqnarray}
for the real part of the order parameter field, and 
\begin{eqnarray}
\stackrel{.}{A}_x^j &=&(1-T)\rho ^j\rho ^{j+x}\sin (-a_xA_x^j-\theta
^j+\theta ^{j+x})\frac 1{a_x}  \label{discreta gauge} \\
&&-\kappa ^2\left[ \frac{A_y^{j+x+y}-A_y^{j+y}-A_y^{j+x}+A_y^j}{a_xa_y}-%
\frac{A_x^{j+y}-2A_x^j+A_x^{j-y}}{a_ya_y}\right]   \nonumber
\end{eqnarray}
for the $x$ component of the gauge field. We choose here to work directly
with the fields, but alternatively the link variables can be used 
\cite{Kato}.

We evolve this equation with a simple Euler scheme, taking time-steps
empirically chosen to be $h={t_0}/{128}$ 
($t_0$ being the time scale defined in eq. \ref{SCALING}), and imposing periodic boundary
conditions. The choice of the time step is very critical because of the very
different and variable time scales involved in this kind of simulation.

We tested grids of $N^2$ sites, with $N=128,256$ and $512$. It is convenient
to employ larger grids, not only because of less granularity in the observed
density, but also because it is possible to achieve sufficient statistics
with fewer runs for ensemble. We have made about 20 runs for each ensemble,
which means that the dispersions of field and defect density are in the $%
\sim 3\%$ level.

We choose the net parameters $a_x=a_y=\xi _0/2$, in order to resolve
adequately the shape of the defects, which are expected to have a final size 
$d\approx \xi _0$.

The initial conditions were set in two different ways.

We obtain a thermal distribution of modes, employing expression (\ref
{Amplitud termica}) as the dispersion of a gaussian distribution of the
mode amplitudes, with $\xi =\xi _0/2,$ that is a temperature equal to five
times $T_c$. The cutoff was set up at the maximum radius in $k$-space, i.e. 
$k_{max}=\frac{2\pi }{\xi _0}$. The results are cutoff independent, in any
case, due to the rapid decay of the short wavelength modes in the first
steps of the quench. At the end of the quench, these modes grow again in
order to define the final shape of the defects. Following \cite{Kato}, we
tested $\mu =10^{-2},10^{-3},10^{-4}$ and $10^{-5} $.

Alternatively, we started the field by choosing uniformly at random 
the phase of the order parameter between $0$ and $2\pi $,
and its modulus between $0$ and $\rho=10^{-4}$ (in one set of simulations)
or between $0$ and $\rho=10^{-6}$ (in another set).
As expected, the results obtained are mostly independent
of how the initial conditions are set \cite{Bray}.

\subsection{Numerical experiments}

The presence of topological defects (or candidate ones) in the field 
$\Delta({\bf x})$, can be determined from
the fact that for any closed curve $C$ we have 
\begin{equation}
\oint\limits_Cd{\bf x.\nabla \theta }({\bf x})=2\pi n_C  \label{circulacion}
\end{equation}
where $\Delta({\bf x})=\rho ({\bf x})\exp [i\theta ({\bf x})]$ and $n_C$ 
is the total topological charge of the defects inside the curve.
By candidate topological defects we mean those who have a net circulation of
the phase, but not the equilibrium profile. That is, we have the phase
defect, but the modulus is still evolving.

The presence of a vortex can be observed trough expression 
(\ref{circulacion}). 
Numerically, we will sum the shortest difference of the phase of the
order parameter field along the lines between nodes of the grid surrounding
each plaquette. Let us call 

\begin{eqnarray}
s(\alpha ,\beta )=\beta -\alpha \nonumber \\
if(s>\pi )\text{ }s=s-2\pi \nonumber \\
if(s<-\pi )\text{ }s=s+2\pi
\end{eqnarray}

So, four neighbor sites $i,j,k,l$; oriented counter clockwise will yield

\begin{equation}
v=\frac 1{2\pi }\left( s(\theta ^j,\theta ^i)+s(\theta ^k,\theta
^j)+s(\theta ^l,\theta ^k)+s(\theta ^i,\theta ^l)\right) =\pm 1,0 
\end{equation}

This device can only measure vorticity $\left| v\right| \leq 1$. So, we can
not detect a pair vortex-antivortex laying in a single plaquette, nor
vortexes with greater vorticity. But this is enough, given the mutual
annihilation of very close vortexes and the almost absolute absence of
higher vorticity, which can be seen in the representation of the phase of
the order parameter field. By the way, periodic boundary conditions provide
a test of the accuracy of the observation, because the net vorticity must
vanish. Higher sensitivity devices can be implemented considering higher
plaquettes, which means more surrounding sites.

The reciprocal representation of the field is obtained trough the usual fast
Fourier transform (FFT) \cite{Numerical Recipes}. In two dimensions this
gives a discrete representation of $\widetilde{\Delta }\left( {\bf k}\right) 
$ at sites ${\bf k}=(n,m)\frac{2\pi }{Na}$ with $n,m=-\frac N2,...,\frac N2%
-1 $, and $a$ is the net parameter supposed equal in both directions. With
our discretization, ${\bf k}=(n,m)\frac{4\pi }{N\xi _0}$, and the domain of
the reciprocal representation embodies the circle $k<\frac{2\pi }{\xi _0}$,
as well as a number of higher modes.

The various mean values of the field can be obtained easily in each time
step, since the space and ensemble average commute. On the other hand, the
power spectrum requires saving each run for further processing.

In order to try to measure the correlation length, we consider the $k^2$
dependence of the ensemble dispersion $g_k^2$ of the amplitude of the $k_{th} $
mode.

\begin{equation}
g_k^2=\left\langle \left| \widetilde{\Delta }\left( {\bf k}\right) \right|
^2\right\rangle 
\end{equation}

For a thermal distribution this is a straight line,

\begin{equation}
\frac 1{g_k^2}=\frac{k^2+\xi ^{-2}}\mu 
\end{equation}

We can estimate $\frac \mu {g_k}\rightarrow \xi ^{-2}$ when $k^2\rightarrow
0 $, through a linear fit of the ensemble media of the spectra. This is a
rough estimate, but gives a qualitative description of the behavior of the
correlation length.

A better determination of the correlation length can be obtained from the
out of equilibrium distribution of modes \cite{Boyanovsky} (see below). At
long wavelengths $k^2<\xi ^{-2}$, 
\begin{equation}
g_k^2\approx he^{-\xi ^2k^2}  \label{correlacion 2}
\end{equation}
This factor can also be measured from a linear fit, and the system quickly
reaches this regime.

\section{Results}

\subsection{Anatomy of a quench}

\begin{figure}[tbp]
\epsfig{file=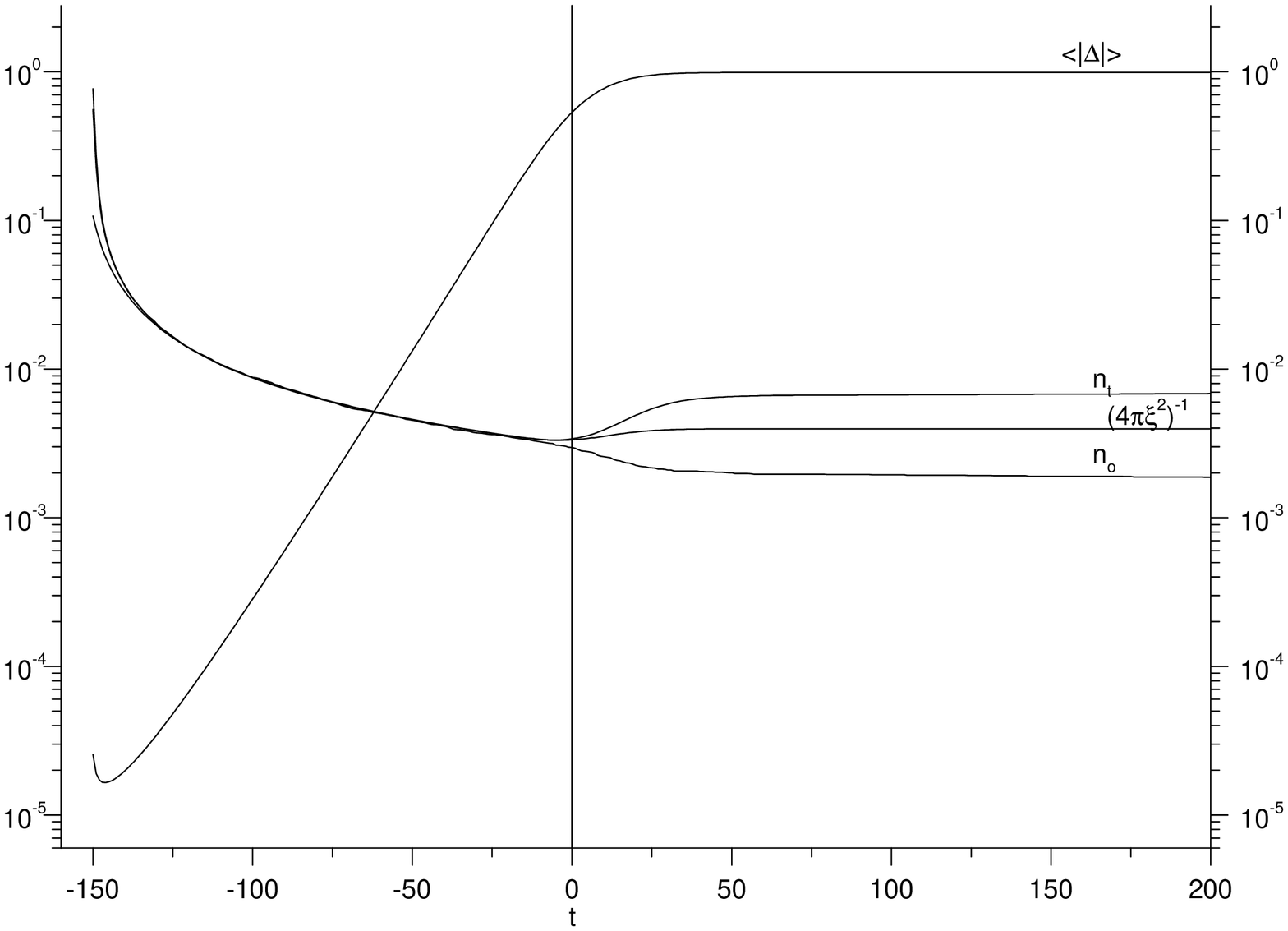,width=11cm,angle=-90}
\caption{
Ensemble average of the absolute value of the order parameter
field $\left\langle \left| \Delta \right| \right\rangle $, the HML
prediction $n_t$ for the defect density, the magnitude $1/\left( 4\pi \xi
^2\right) $ (where $\xi $ is the correlation length measured form a fit of
the long wavelength part of the correlation function, as in eq. (\ref
{correlacion 2})) and the observed defect density $n_o$, all as functions of
time. 
This run corresponds to $\mu =10^{-4}$ and $T=2.$. Initially $\xi $
differs from both the predicted and observed densities, but the rise of the
gauge field smears out this difference.
While the graph of $\left\langle \left| \Delta \right| \right\rangle $ 
changes concavity at $t=0$, this is not easily appreciated due to the 
distortion caused by the linear-logarithmic scales.
The same curve is plotted in linear-linear scales in fig. 
[\ref{Fig cocientes}]
}
\label{Fig tipica}
\end{figure}

Figure \ref{Fig tipica} shows a typical evolution of a quench. We have
plotted the ensemble average of the absolute value of the order parameter
field $\left\langle \left| \Delta \right| \right\rangle $, the HML
prediction $n_t$ for the defect density, the magnitude $1/\left( 4\pi \xi
^2\right) $ (where $\xi $ is the correlation length measured form a fit of
the long wavelength part of the correlation function, as in eq. (\ref
{correlacion 2})) and the observed defect density $n_o$, all as functions of
time. This run corresponds to $\mu =10^{-4}$ and $T=2.$ We have chosen $t=0$
as the point where the second derivative of the order parameter changes sign.

\begin{figure}[tbp]
\epsfig{file=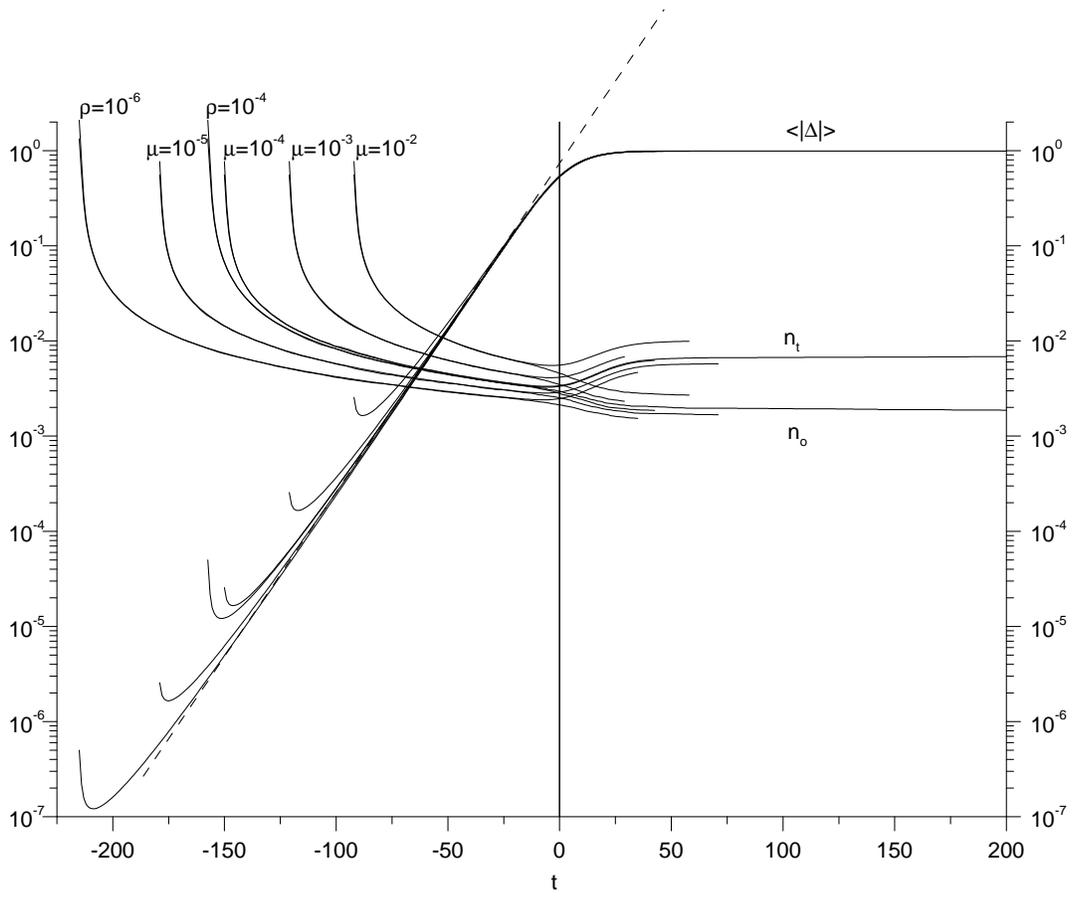,width=11cm,angle=-90}
\caption{$\left\langle \left| \Delta \right| \right\rangle ,$ $n_t$ and $n_o$
for all the six ensembles. All ensembles has been synchronized at the
inflection points of the curves $\left\langle \left| \Delta \right|
\right\rangle $. Also represented are the defect densities predicted and
observed.}
\label{Fig compare}
\end{figure}

The Figure \ref{Fig compare} shows 
$\left\langle \left| \Delta \right|\right\rangle ,$ $n_t$ 
and $n_o$ for all the six ensembles tested. 
The time scales are shifted in order to make all the inflection points of 
$\left\langle\left| \Delta \right| \right\rangle $ to coincide. 
We can see that the behavior of each ensemble is essentially the same.

Once the simulation begins, the
gauge field (initially null) adjusts itself in order to follow the order 
parameter field, reacting back on it.
This is what can be expected from the very different time scales involved in
the equations (\ref{discreta orden}) and (\ref{discreta gauge}), and it
appears in the graph as the initial decay of the order parameter field. 
The observed discrepancy between predicted and observed defect densities is
quickly smeared out by the evolution of the field. Actually neither of them
is reliable this early in the simulation, since the prescribed thermal
distribution has too much power at short wavelengths, which can be
eliminated with a cutoff. 
Furthermore the algorithm to identify defects is blind to higher vorticity 
monopoles. 
Both errors are quickly self-corrected, though, as the density of defects 
decays and the distribution of modes becomes a gaussian function, 
as in eq. (\ref{correlacion 2}). 
This gaussianity should not be confused with that of the ensemble, but just
refers to the shape of $g(k)^2$.

\begin{figure}[tbp]
\epsfig{file=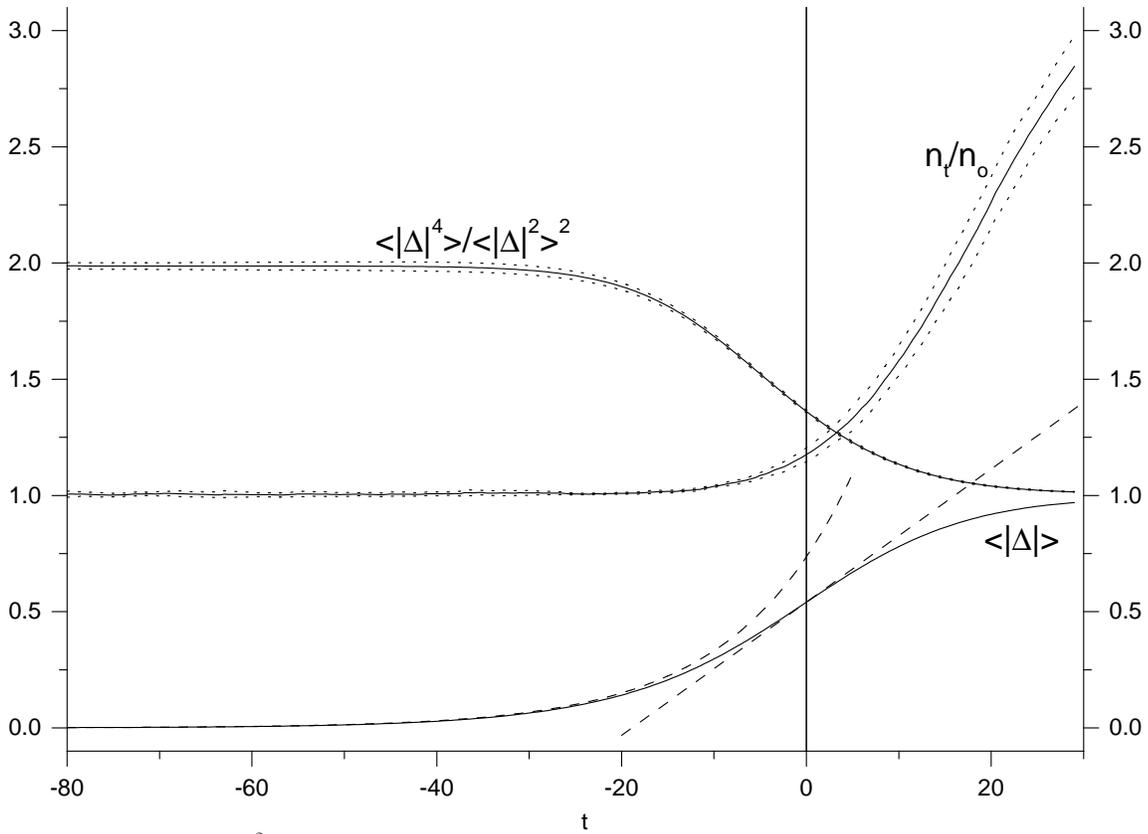,width=11cm,angle=-90}
\caption{$\left\langle \left| \Delta \right| ^4\right\rangle \left\langle
\left| \Delta \right| ^2\right\rangle ^{-2}$, as a function of time,
together with the ratio $n_t/n_o$ and $\left\langle \left| \Delta \right|
\right\rangle .$ The plot represents the average of these quantities over
all six ensembles; the dot lines around the first two represent the
dispersion between ensembles. The dashed lines around the plot of the order
parameter represent the empirical fit to an exponential, and the tangent at
the inflection point. The ratios show that the predicted and observed defect
densities agree very well even when the field distribution ceases to be
gaussian.}
\label{Fig cocientes}
\end{figure}

The first stage of the quench (once the transient is over) is characterized 
by the exponential grow of the order parameter field, which can be parametrized 
empirically as 
$\left\langle \left| \Delta \right| \right\rangle =%
2e^{0.159t-1}\approx 2e^{\frac t{2\pi }-1}$ 
(the dashed line in Figure \ref{Fig compare}), where $t$ is the 
synchronized time. 
While this is to be expected from the growth of the
spinodal instability, it must be observed that we are already beyond the
linear regime at this stage. 
This is the regime where the HML prediction is essentially exact.
The ensemble probability density of the field is clearly gaussian, as can be
tested by the ratio $\left\langle \left| \Delta \right| ^4\right\rangle
\left\langle \left| \Delta \right| ^2\right\rangle ^{-2}$. In fact, for a
gaussian ensemble, considering only diagonal terms of the correlation we have 
$\left\langle \left| \Delta \right| ^4\right\rangle =%
2\left\langle \left| \Delta \right| ^2\right\rangle ^{2}$ .
We have plotted this ratio vs time, in Figure \ref{Fig cocientes}, 
together with the ratio $n_t/n_o$ and 
$\left\langle \left| \Delta \right| \right\rangle .$ 
The plot
represents the average of these quantities over all six ensembles; the dot
lines around the first two represent the dispersion between ensembles. The
dashed lines around the plot of the order parameter represent the empirical
fit to an exponential, and the tangent at the inflection point.

\begin{figure}[tbp]
\epsfig{file=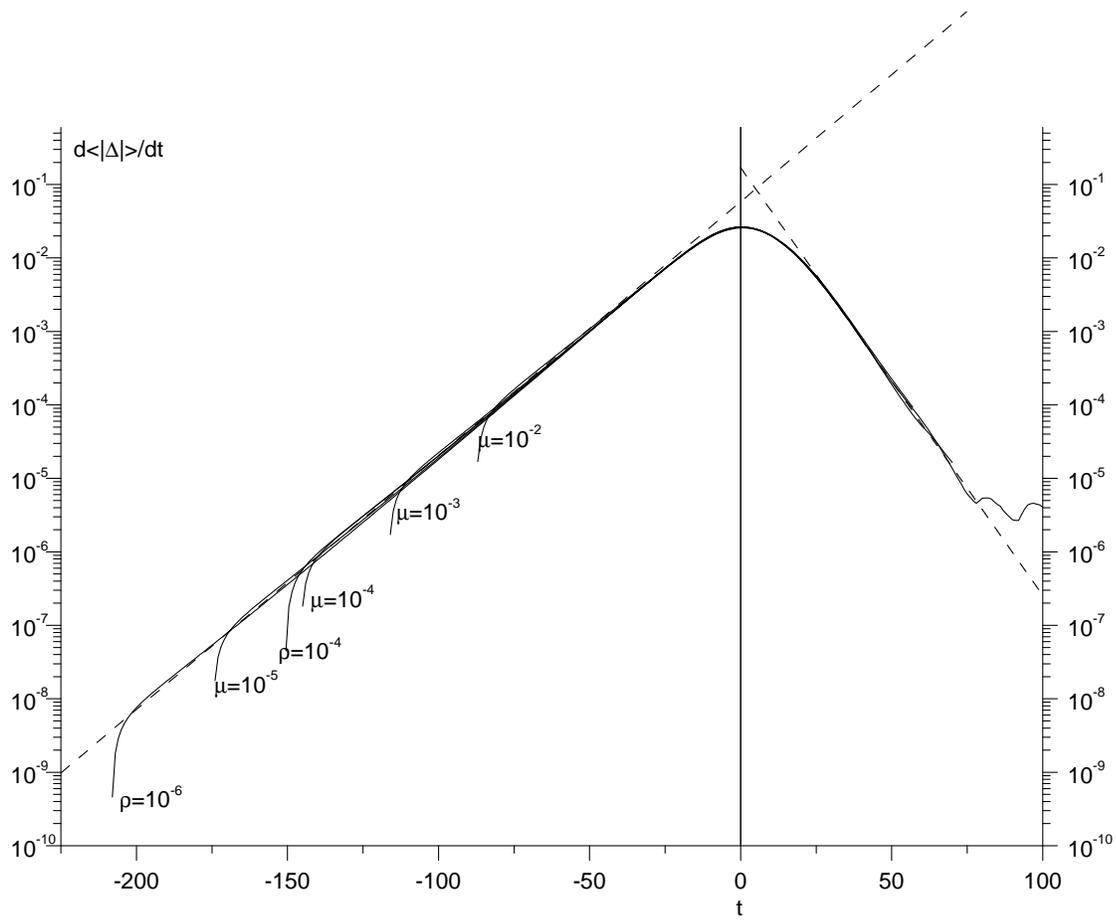,width=11cm,angle=-90}
\caption{Time derivative of $\left\langle \left| \Delta \right|
\right\rangle $, there is no symmetry at all between both sides of the
inflexion point.}
\label{Fig criticos}
\end{figure}

When the growth of the average order parameter ceases to be exponential 
(around $t\sim -50$, see fig. \ref{Fig criticos}), we enter a transition 
regime where first the gaussianity of the ensemble, and then the HML 
prediction cease to hold

In the final stage, from $t\sim 25$ on, the approach to the equilibrium value 
is also exponential, as can be appreciated in Figure \ref{Fig criticos}, 
where we have plotted the time derivative of 
$\left\langle \left| \Delta \right|\right\rangle$, 
with linear-logarithmic scales. 
In this final stage, the exponential behavior of the field is 
$\propto 1-e^{-0.134t}$, and the topological defects have attained almost 
their stable profiles (see below).

We can see that, as long as the ensemble is Gaussian, the HML prediction is
exact for all practical purposes. Around $t\sim -30$, both gaussianity and
the exponential growth of the order parameter break down; however, the HML
estimate is still a good approximation until later times, $t\sim -10$.

\subsection{Evolution of the structure function}

The two main regimes in the evolution of the quench, the early one dominated
by the growth of the order parameter, and the late one dominated by the
evolution of the defect network, are also clearly seen in the evolution of
the structure function, namely the Fourier transform of the equal time order
parameter correlation function (for the structure function in systems with
global symmetry, see ref. \cite{globsym}).

At early times and long wavelengths, the order parameter and the gauge field
are essentially decoupled. Under this approximation, the dynamical equation
eq. (\ref{TDGL-adim}) becomes,

\begin{equation}
\frac \partial {\partial t}\Delta =\frac 1{12}\left[ {\bf \nabla }^2\Delta
+\Delta \right]  \label{TDGL-et}
\end{equation}

Assuming that each mode has a random initial phase, the theoretical
prediction for the structure function at early times and long wavelengths is

\begin{equation}
g_k^2\sim {\rm exp}\left\{ \left( \frac t6\right) \left[ 1-k^2\right]
\right\}  \label{sf-et-lw}
\end{equation}

\begin{figure}[tbp]
\epsfig{file=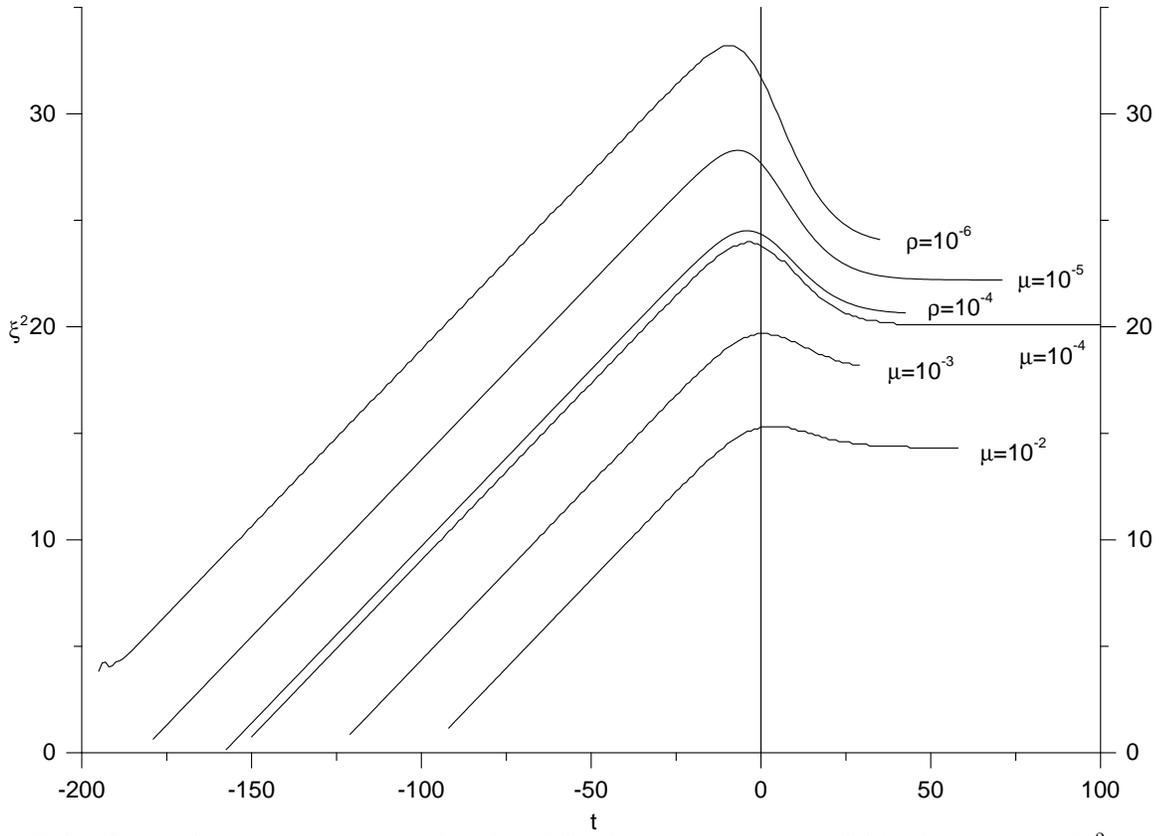,width=11cm,angle=-90}
\caption{ Correlation length squared as a function of time for each ensemble
The initial linear growth of $\xi ^2$ stops simultaneously with the slow
down in the growth of the order parameter field. These curves agree very
well when represented in simulation time, rather than shifting time to make
the inflection points to coincide.}
\label{Fig chiqu}
\end{figure}

In particular, the correlation length, determined from the scaling condition 
$g_k^2\sim f\left( \xi k\right) $, grows as $\sqrt{t}$. In 
Figure \ref{Fig chiqu} 
we have plotted the correlation length squared as a function of
time for each ensemble; the result clearly agree with expectations. In this
regime the calculation (\ref{integral kcu}) reduces to 
$n_t\approx \frac{1}{4\pi\xi ^2}$, 
which, as we have seen, agrees very well with the observed density.

For later times and wavelengths shorter than the average defect separation,
the structure function is dominated by the profile of an isolated defect.
The Abrikosov-Gorkov vortex centrated at the origin is given by 
\begin{equation}
\Delta \sim \left( 1-e^{-\frac r{r_1}}\right) e^{i\theta }  \label{Abrikosov}
\end{equation}
where $r_1$ is the characteristic size of the defect. The Fourier transform
of this shape gives 
\begin{equation}
\widetilde{\Delta }_k\sim \frac{2\pi }{k^2}\left[ 1-\left(
1+r_1^{-2}k^{-2}\right) ^{-\frac 32}\right] e^{-i{\bf k\cdot x}_0}
\label{abri-ft}
\end{equation}

With $x_0$ the position of the defect.
Considering short wavelengths, as compared to the average defect separation,
we obtain 
\begin{equation}
g_k^2\sim \left| \widetilde{\Delta }_k\right| ^2n_o  \label{Gorkov}
\end{equation}
where $n_o$ is the observed density of defects. By construction, a grid can
not support singularities like that at the origin in (\ref{Abrikosov}), but
the power law characteristic of (\ref{Gorkov}) is the kind of spectrum we hope
to find in the final regime of the quench.

\begin{figure}[tbp]
\epsfig{file=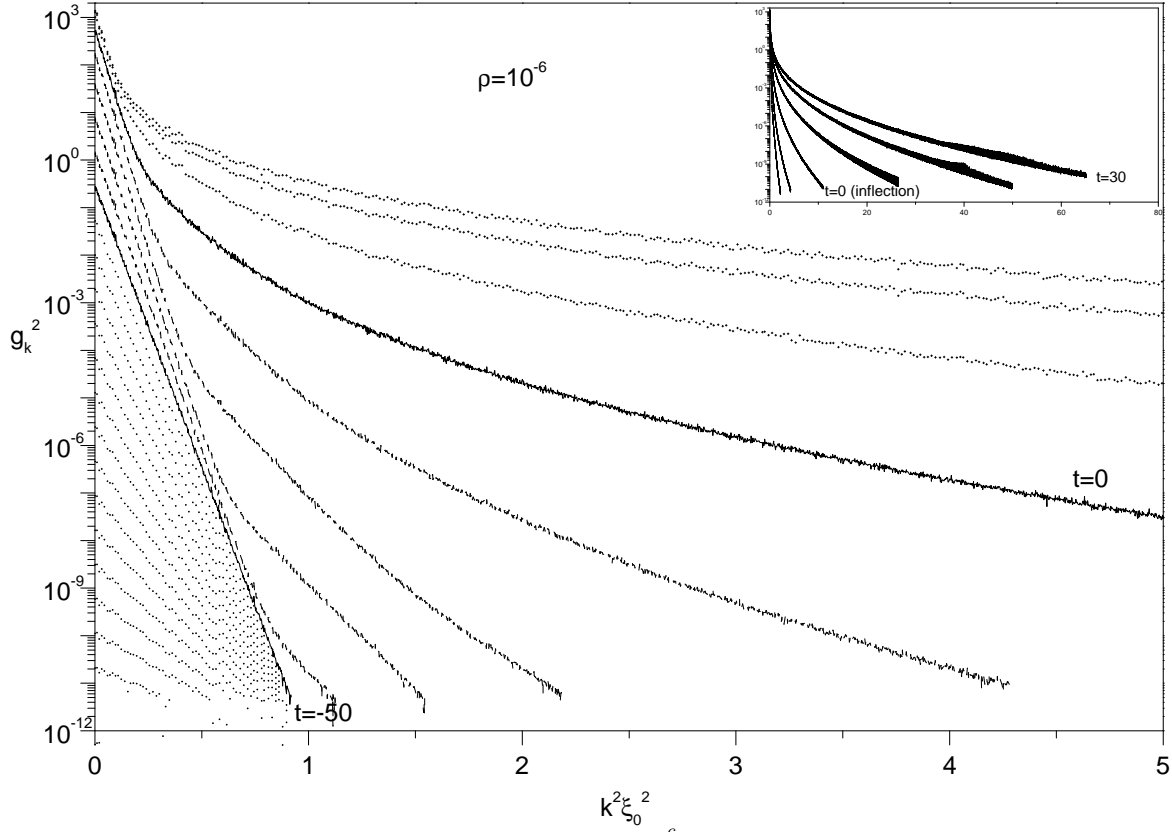,width=11cm,angle=-90}
\caption{ Power spectra measured for the ensemble $\rho =10^{-6}$ at equally
spaced different times, as a function of $\xi _0^2k^2$, for random initial
conditions and $\rho =10^{-6}$; the vertical scale is logarithmic. The first
bold curve corresponds to the time at which the spectrum leaves the behavior 
$g_k^2\approx he^{-\xi ^2(k^2-1)}$ . The second corresponds to the
inflection time. The insert shows the growth of short wavelength modes
needed for the final shape of the defects. }
\label{Fig Espectro 1}
\end{figure}

\begin{figure}[tbp]
\epsfig{file=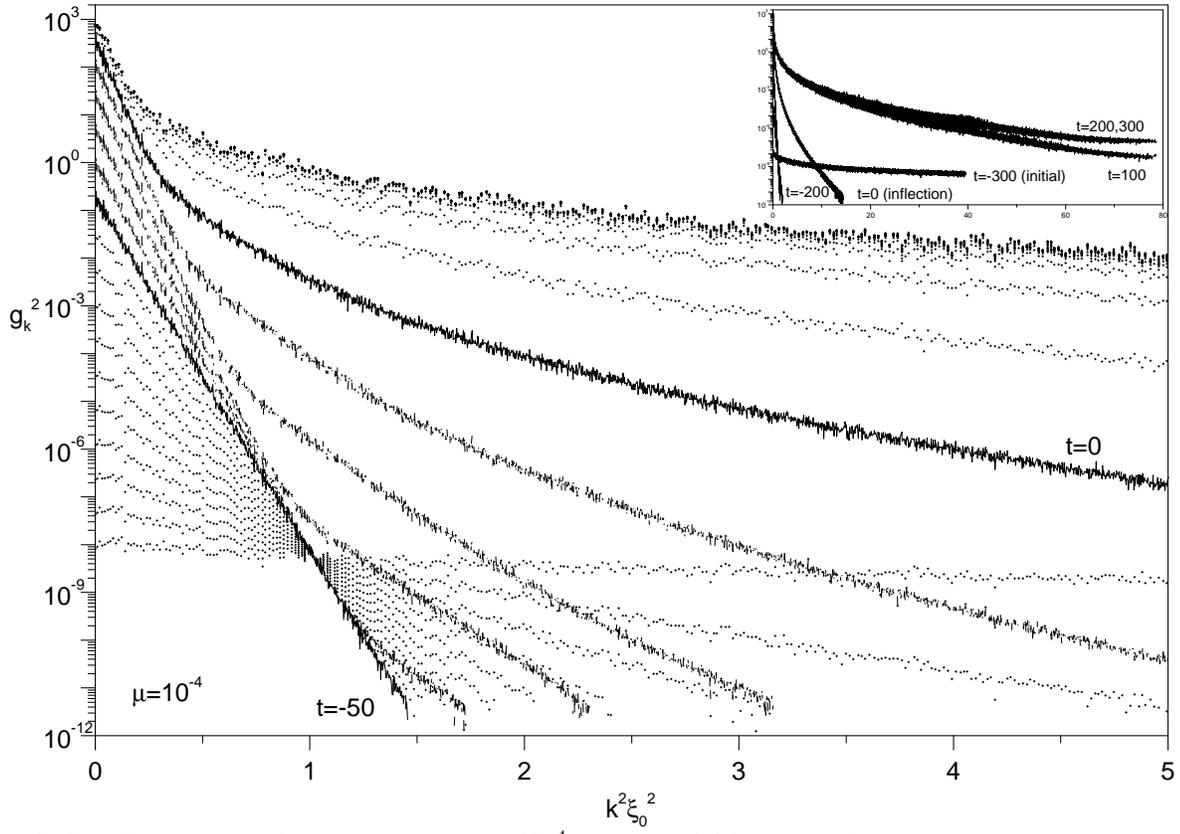,width=11cm,angle=-90}
\caption{ Power spectra for the ensemble $\mu =10^{-4}$. Note the initial
decay of the short wavelength modes. At late times these modes grow again
and for this ensemble the shape of the vortex almost reaches its
equilibrium form.}
\label{Fig Espectro 2}
\end{figure}

Figure \ref{Fig Espectro 1} displays the evolution of the structure function
(plotted every 10 time units) as a function of $\xi _0^2k^2$, for random
initial conditions and $\rho =10^{-6}$; the vertical scale is logarithmic.
The early plots clearly display the gaussian behavior predicted by eq. (\ref
{sf-et-lw}). After $t=-50$ (bold line) the short wavelength modes begin to
grow beyond the early times prediction. The second bold line represents the
structure function at $t\sim 0$, that is, the beginning of the late times
regime. The insert shows the same plot, for a wider range in wavenumber.
Figure \ref{Fig Espectro 2} shows the same for the thermal initial
conditions, and $\mu =10^{-4}$. The same behavior is obtained, being
remarkable that the value of $g_k^2$ for $k\xi _0=1$ remains constant over
the early times regime.

\begin{figure}[tbp]
\epsfig{file=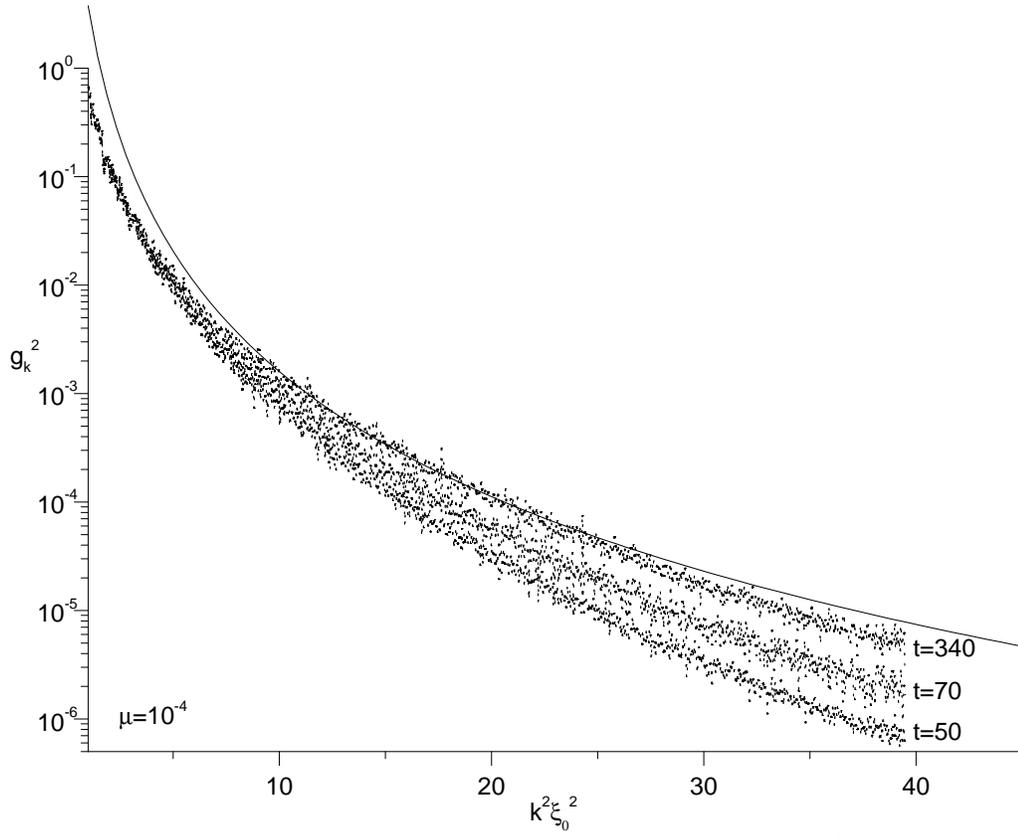,width=11cm,angle=-90}
\caption{ Detail of the previous figure. The solid curve correspond to $%
\gamma \left( x^{-1}\left( 1-\left( 1+x\right) ^{-\frac 32}\right) \right)
^2 $ (see eq. (\ref{abri-ft})) , the factor $\gamma $ is approximately equal
to the final density of defects. Only the modes corresponding to the maximum
circle in reciprocal space are depicted ($k^2\xi _0^2<4\pi ^2$).}
\label{Fig Abrikosov}
\end{figure}

In Figure \ref{Fig Abrikosov} we have contrasted three of the structure
functions shown in Figure \ref{Fig Espectro 2} (corresponding to times $t=50,$ 
$70$ and $340$) with the structure function of an isolated defect, as given
by eqs. (\ref{abri-ft}) and (\ref{Gorkov}), given by the solid line. For
visual effect, we have overestimated slightly the value of $n_o$ in eq. (\ref
{Gorkov}). We have set $r_1=1$, as predicted by theory.

\subsection{The epoch of defect formation}

Candidate defects exist from the very beginning of the quench, as can be
detected from the phase of the order parameter field, and predicted by the
HML formula. It is the development of a well defined vortex and the pattern
of supercurrents around it which makes the system leave the gaussian
distribution. The exponential growth slows down as soon as the modes stop
behaving almost independently ($t\sim -30$). The formation of the shape of
the vortex itself is also starting at this time.

We may follow the formation of the vortexes through the evolution of the
''kinetic '' free energy \cite{Landau} 
\begin{equation}
K=\frac 1V\int d^2x\;\left| \left( -i\nabla -{\bf A}\right) \Delta \right|
^2=\frac 1V\int d^2x\;\left\{ \left| \nabla \rho \right| ^2+\rho ^2\left|
\nabla \theta -{\bf A}\right| ^2\right\}  \label{energya}
\end{equation}

where $\Delta =\rho e^{i\theta },$ and the last term corresponds to the
supercurrents. Initially $K$ is very low, and starts building up with the
steeping of the field gradients around the candidate defects. When the
defect attains its final shape, both $\nabla \rho $ and the current die out
outside the core, but there is a core contribution left, and so $K$ reaches
a final value which is proportional to the defect density. The subsequent
evolution of $K$ simply follows the slow decay of the defect density due to
defect - defect annihilation.

\begin{figure}[tbp]
\epsfig{file=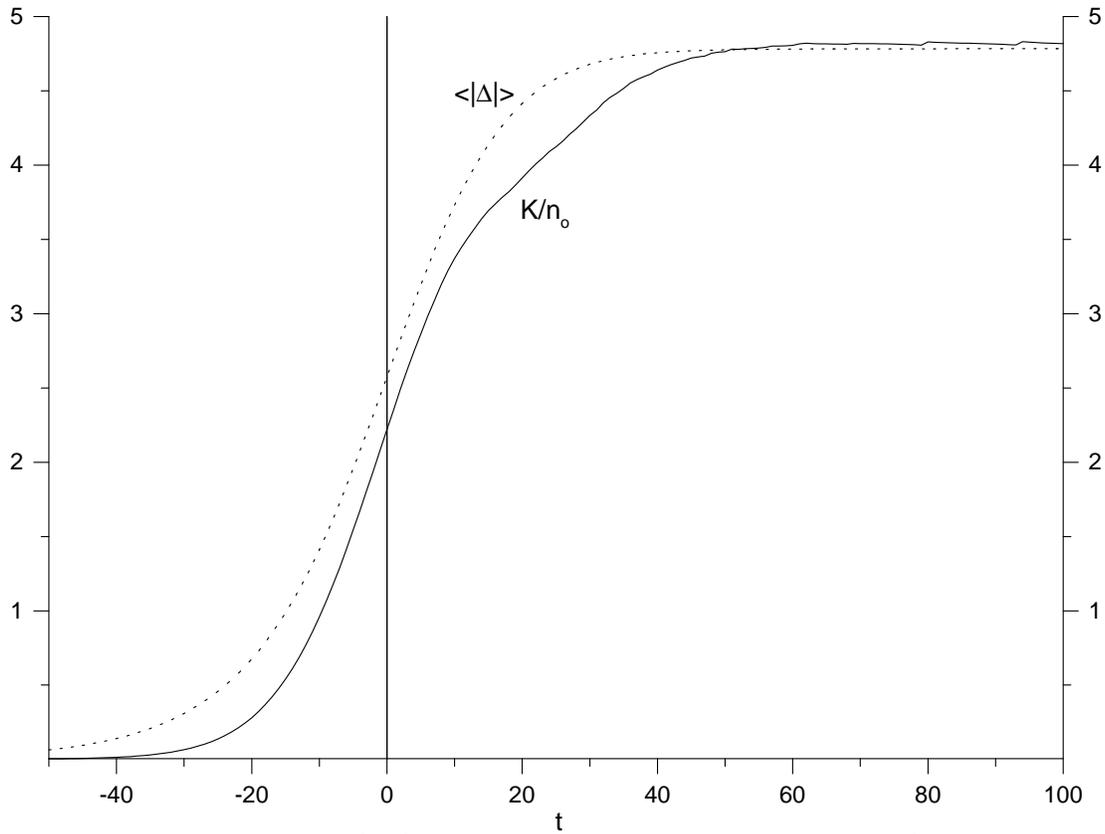,width=11cm,angle=-90}
\caption{ Ensemble average of $K/n_o$ (also averaged over the six ensembles
considered) as a function of time. For comparison purposes we also show $%
\gamma \left\langle \left| \Delta \right| \right\rangle $ (dashed line),
where $\gamma =4.82$ is the asymptotic value of $K/n_o$, showing the final 
between the kinetic term and the defect density.}
\label{Fig Kinefinal}
\end{figure}

The Figure \ref{Fig Kinefinal} shows the ensemble average of $K/n_o$ (also
averaged over the six ensembles considered) as a function of time. For
comparison purposes we also show $\gamma \left\langle \left| \Delta \right|
\right\rangle $ (dashed line), where $\gamma =4.82$ is the asymptotic value
of $K/n_o$. At time $t\sim 70$ the final shape and current has been reached,
and the tiny fluctuations are due to transients corresponding to vortex
annihilation.

A remarkable implication of Figure \ref{Fig Kinefinal} is that defect formation
occurs at a relatively well defined epoch, from $t\sim -30$ to $40$.

\subsection{Final Remarks}

This paper attempts to answer the question of how reliable are estimates of
the defect density based on the approximation of gaussian ensembles as
applied to nonlinear phase transitions. To this effect, we have measured
independently the gaussian prediction and the actual defect density as a
function of time after an instantaneous quench to zero temperature in a two
dimensional superconductor. The evolution of the quench goes through three
stages, an initial one dominated by the unfolding of the spinodal
instability, a final one dominated by defect - defect interactions, and a
transitional stage when most defects are actually formed.

We find that the gaussian estimate is essentially exact over the early
regime, and continues to be accurate almost to the end of the epoch of
defect formation. This moment in time is marked by the jump in the ''kinetic
''energy $K$ (see eq. (\ref{energya})); it is also the time when the order
parameter reaches about half of its equilibrium value. While the gaussian
estimate itself is not reliable beyond this point, it is still a valid tool
to fix the initial conditions of the defect network, whose subsequent
evolution must be investigated by other means (like those in this paper, or
in ref. \cite{Vilenkin}).

These results confirm theoretical expectations, but it is nevertheless
satisfactory to have solid numerical proof of formerly theoretical
conjectures. The very detailed view of the process of defect formation which
is afforded by our simulations should also be valuable in investigating more
subtle processes, such as preheating during the non equilibrium phase
transition \cite{preh} or instabilities due to strong field effects \cite
{tsunami}. Also, by doing a more complete simulation, where we could also
control the quench rate, it ought to be possible to investigate Zurek's
conjecture about the scaling of the defect density with the quench rate 
\cite{Zurek,Yates-Zurek,He4}. 
Finally, it is of interest to perform simulations in regions of
parameter space approaching actual experimental contexts. We continue our
research in these manyfold directions.

\section{Acknowledgments}

It is a pleasure to acknowledge many conversations with Bei-lok Hu, Steve
Ramsey and Greg Stephens.

This work is partially supported by UBA, CONICET, Fundacion Antorchas and
FOMEC.


\end{document}